\documentclass[aps,prl,showpacs,notitlepage,nofootinbib,superscriptaddress,floatfix,showkeys,twocolumn]{revtex4-2}

\usepackage[T1]{fontenc} 
\usepackage[utf8]{inputenc}
\usepackage{pstool}
\usepackage{hyperref}
\usepackage{amsmath,amssymb}
\usepackage{float}
\usepackage{graphicx}
\usepackage{graphics}
\usepackage{bm}
\usepackage{latexsym}
\usepackage{psfrag}
\usepackage[dvipsnames]{xcolor}
\usepackage{subfigure}
\usepackage[normalem]{ulem}
\usepackage{textcomp}
\usepackage{comment}
\usepackage{braket}
\usepackage{mathalpha}
\usepackage{tabularx}
\usepackage{placeins}   

\def\bea{\begin{eqnarray}}
\def\eea{\end{eqnarray}}
\def\nn{\nonumber}

\def\f{\frac}

\def\d{{\rm d}}
\def\Mpl{M_{_{\mathrm{Pl}}}}
\def\mpcinv{{\rm Mpc}^{-1}}

\def\vk{{\bm{k}}}
\def\vq{{\bm{q}}}
\def\hvk{{\hat{\bm{k}}}}

\def\cR{{\cal R}}

\def\ps{\mathcal{P}_{_{\mathrm{S}}}}

\def\ph{\mathcal{P}_{h}}
\def\barpow1d{\mathcal{P}^{1\rm D}_{\mathrm{b}}}

\def\ns{n_{_{\mathrm{S}}}}
\def\As{A_{_{\rm S}}}

\def\ogw{\Omega_{_{\mathrm{GW}}}}
\def\fnl{f_{_{\rm NL}}}
\def\gnl{\tau_{_{\rm NL}}}
\def\tgnl{\tilde{\tau}_{_{\rm NL}}}

\allowdisplaybreaks

\begin{document}

\title{{\it Twisted echoes of an odd quartet:}\\
Scalar-induced gravitational waves as a probe of primordial parity-violation}
\author{H.~V.~Ragavendra}
\email{ragavendra.hv@pd.infn.it}
\affiliation{Dipartimento di Fisica e Astronomia “Galileo Galilei”, Universit\`{a} degli Studi di Padova, Via Marzolo 8, I-35131 Padova, Italy}
\affiliation{Istituto Nazionale di Fisica Nucleare (INFN), Sezione di Padova, Via Marzolo 8, I-35131 Padova, Italy}
\author{Nicola Bartolo}
\email{nicola.bartolo@pd.infn.it}
\affiliation{Dipartimento di Fisica e Astronomia “Galileo Galilei”, Universit\`{a} degli Studi di Padova, Via Marzolo 8, I-35131 Padova, Italy}
\affiliation{Istituto Nazionale di Fisica Nucleare (INFN), Sezione di Padova, Via Marzolo 8, I-35131 Padova, Italy}
\affiliation{Istituto Nazionale di Astrofisica (INAF), Osservatorio Astronomico di Padova, vicolo dell’ Osservatorio 5, I-35122 Padova, Italy}

\begin{abstract}
Parity-violation leaves tell-tale trails in many cosmological observables. We illustrate parity-odd primordial scalar trispectra, that despite being of modest strength, impart detectable chirality to scalar-induced gravitational waves (SIGW). This allows us to impose strong bounds on the parity-odd part of trispectrum. Over certain scales, we find SIGW directly quantify parity-violation in primordial non-Gaussianity, unobscured by the Gaussian contribution. Our results call for treatment of SIGW and parity-odd trispectrum as complementary predictions of parity-violating theories.
\end{abstract}

\maketitle

{\it Introduction}---The phenomenon of scalar-perturbations inducing tensor 
perturbations at the second order of perturbation theory has been long examined 
in the literature~\cite{Tomita:1967wkp,Matarrese:1992rp,Matarrese:1993zf,Matarrese:1996pp,
Matarrese:1997ay,Ananda:2006af,Baumann:2007zm,Bartolo:2007vp,Guzzetti:2016mkm,Kohri:2018awv}\,
(see~\cite{Domenech:2021ztg} for a review).
Recently, these scalar-induced gravitational waves (SIGW) have been considered
as a potential candidate to explain the signature of stochastic GW claimed by 
the Pulsar Timing Array (PTA) collaboration~\cite{NANOGrav:2023hvm,Cai:2023dls,
Franciolini:2023pbf,Chang:2023vjk,Chang:2023aba,Li:2023xtl,Firouzjahi:2023lzg,
Inui:2023qsd,Yi:2023tdk,Liu:2023ymk,Domenech:2024rks,Maity:2024odg,Wang:2025kbj}.
There have also been efforts in the literature to examine the generation and 
direct interferometric detection of chirality in stochastic
GW~\cite{Satoh:2007gn,Takahashi:2009wc,Sorbo:2011rz,Anber:2012du,Caprini:2014mja,Bielefeld:2014nza,Bartolo:2014hwa,Anber:2016yqr,Cai:2016ihp,Thorne:2017jft,Okano:2020uyr,Watanabe:2020ctz,Okano:2020uyr,RoperPol:2021xnd,Brandenburg:2021pdv,Brandenburg:2021bfx,Cai:2021uup,Bastero-Gil:2022fme,Bari:2023rcw,Dimastrogiovanni:2023oid,Garcia-Saenz:2023zue,Alexander:2024klf,Zhang:2024vfw,Feng:2024yic,Alexander:2024klf,Ozsoy:2024apn,
Crowder:2012ik,Kato:2015bye,Smith:2016jqs,Garcia-Bellido:2016dkw,Qin:2018yhy,Inomata:2018rin,Domcke:2019zls,Belgacem:2020nda,Orlando:2020oko,Sato-Polito:2021efu,Cruz:2024esk,Chen:2024ikn}.

Violation of the symmetry of parity in the universe is a compelling phenomenon with 
several cosmological implications, such as cosmic 
birefringence~\cite{Minami:2020odp,Diego-Palazuelos:2022dsq,Komatsu:2022nvu,
Greco:2022xwj,Arcari:2024nhw,Gruppuso:2025ywx}.
The presence of chirality in cosmological GW shall be a signature of parity-violation
in the universe.
Another signature shall be in the N-point correlation functions of the perturbations, 
with the lowest order (for scalar perturbations) sensitive to parity being the 
irreducible four-point correlation, the trispectrum $\cal T$.
There have been recent attempts searching for the parity-odd component of the 
trispectrum in the dataset of Baryon Oscillation Spectroscopic Survey 
(BOSS)~\cite{Hou:2022wfj,Philcox:2022hkh,Krolewski:2024paz,Hewson:2024rnb}.

In the presence of a non-negligible trispectrum $\cal T$, it is logical to expect 
its signatures in sourcing SIGW.
Moreover, if there is a parity-odd component to the trispectrum ${\cal T}_{\rm odd}$, 
the signature of parity-violation can be inherited by the SIGW, thereby acquiring a 
non-vanishing chirality~\cite{Garcia-Saenz:2023zue}.
In this analysis, we examine such an acquisition of chirality by SIGW 
from ${\cal T}_{\rm odd}$.
We quantify the chirality by computing the ratio of spectral densities of the degree of
circular polarization $V$ to the intensity $I$ of the scalar-induced tensor perturbations
using certain generic templates of trispectrum.
The templates of trispectrum we employ are fair representatives of the models of
inflation violating parity and enable us to relate the induced chirality to the
behavior of ${\cal T}_{\rm odd}$.
We arrive at theoretical bound on the amplitude of ${\cal T}_{\rm odd}$, 
utilizing the inequality relating the Stokes parameters $V$ and intensity $I$
of the GW.
We also identify interesting regimes of wavenumbers, in cases with peaked
forms of scalar power- and tri-spectra, where the chirality is determined by the ratio of
parity-odd to parity-even parts of $\cal T$.

This generic mechanism shall have interesting implications for different classes of parity-violating scenarios.
For instance, models with gravitational Cherns-Simons coupling predict a highly suppressed chirality for primary tensor perturbations (for example~\cite{Satoh:2010ep,Bartolo:2017szm,Bartolo:2018elp,Qiao:2019hkz,Bartolo:2020gsh,Creque-Sarbinowski:2023wmb,Moretti:2024fzb,Orlando:2025fec}), 
but could lead to substantial amplitude of ${\cal T}_{\rm odd}$~\cite{Creque-Sarbinowski:2023wmb,Moretti:2024fzb,Orlando:2025fec}. 
Hence SIGW shall play a dominant role in determining the chirality of their total system of GW.
On the other hand, models with coupling between axion and gauge fields predict their 
primary GW to be maximally chiral 
(for instance~\cite{Sorbo:2011rz,Adshead:2013qp,Caprini:2014mja,Dimastrogiovanni:2016fuu,
Dimastrogiovanni:2025snj}).
In such models, the chirality of the total GW shall acquire new contributions from SIGW.

Regarding notations, we denote the conformal time as $\eta$, the reduced Planck mass 
as $\Mpl = (8\pi G)^{-1/2}$ and set $\hbar = c =1$ throughout the manuscript.


{\it Role of trispectrum in SIGW}---The two-point correlation of the 
scalar-induced tensor perturbations in Fourier space $h_\vk^\lambda$, is 
given by~\cite{Kohri:2018awv,Adshead:2021hnm,Ragavendra:2021qdu,Perna:2024ehx}
\begin{eqnarray}
\braket{h^{\lambda_1}_{\vk_1}h^{\lambda_2}_{\vk_2}} &=& \f{8}{k_1k_2\eta^2}\,
\iint \f{\d^3 \vq_1\d^3 \vq_2}{(2\pi)^6}\,
Q^{\lambda_1}(\vk_1,\vq_1) Q^{\lambda_2}(\vk_2,\vq_2) \nn \\ 
& & \tilde I_2\left(\f{q_1}{k_1},\f{\vert \vk - \vq_1 \vert}{k_1},
\f{q_2}{k_2},\f{\vert \vk_2 - \vq_2 \vert}{k_2}\right)\nn \\
& & \times \braket{\cR_{\vq_1}\cR_{\vk_1-\vq_1}\cR_{\vq_2}\cR_{\vk_2-\vq_2}}\,.
\label{eq:h1h2}
\end{eqnarray}
The function $\tilde I_2$ captures the factors arising from the transfer functions
relating the Bardeen potential to the primordial scalar perturbation $\cR_\vk$ along with the
Green's function associated with the inhomogeneous solution of $h_\vk^\lambda$.
The factor of $Q^\lambda$ for a given polarization $\lambda$ arises from 
the contraction of the polarization tensor with the momenta of integration 
(cf. appendix).
The scalar four-point function sourcing the two-point function of $h^\lambda_\vk$ shall 
have a component reducible to the scalar power spectrum $\ps$ and an irreducible 
component, the trispectrum $\cal T$, which is defined by the relation
\begin{eqnarray}
\langle \cR_{\vk_1}\cR_{\vk_2}\cR_{\vk_3}\cR_{\vk_4}\rangle_c = 
(2\pi)^3\,{\cal T}(\vk_1,\vk_2,\vk_3,\vk_4) \nn \\\
\times \delta^{(3)}(\vk_1+\vk_2+\vk_3+\vk_4)\,.
\end{eqnarray}
The  dimensionless power spectrum associated with $h^\lambda_\vk$, $\ph^{\lambda_1\lambda_2}$ 
can then be computed as
\begin{eqnarray}
\braket{h^{\lambda_1}_{\vk}h^{\lambda_2}_{\vk'}} &=& (2\pi)^3
\f{2\pi^2}{k^3}\ph^{\lambda_1\lambda_2}(k,\eta)\,\delta^{(3)}(\vk+\vk')\,,
\end{eqnarray}
which shall be related to $\ps$ and $\cal T$ as
\begin{eqnarray}
\ph^{\lambda_1\lambda_2}(k,\eta) &=& \f{4}{\pi}\f{k^3}{(k\eta)^2}\,
\bigg[ \int \d^3 \vq\,Q^{\lambda_1}(\vk,\vq) Q^{\lambda_2}(-\vk,-\vq) \nn \\ 
& & \times \tilde I_1\left(\f{q}{k},\f{\vert \vk - \vq \vert}{k}\right)
\f{\ps(q)\ps(\vert \vk-\vq \vert)}{q^3\vert \vk-\vq \vert^3} \nn \\
& & +\,\f{1}{\pi}\iint \f{\d^3 \vq_1 \d^3 \vq_2}{(2\pi)^6} \, Q^{\lambda_1}(\vk,\vq_1) 
Q^{\lambda_2}(-\vk,-\vq_2) \nn \\
& & \times \tilde I_2\left(\f{q_1}{k},\f{\vert \vk - \vq_1\vert}{k},
\f{q_2}{k},\f{\vert \vk - \vq_2\vert}{k}\right) \nn \\
& & \times {\cal T}(\vq_1,\vk-\vq_1,-\vq_2,\vq_2-\vk)\bigg]\,.
\label{eq:ph-l1-l2}
\end{eqnarray}
We have retained the arbitrary combination of $\lambda_1$ and $\lambda_2$ in 
defining $\ph$.
If we inspect the possible combinations of $\lambda_1,\lambda_2$ in the above 
expression, we shall find that only terms with $\lambda_1=\lambda_2$ survive 
the integration, essentially due to the statistical homogeneity and isotropy of 
the perturbations. 

We may identify $\ph$ associated with the intensity $I$ of the correlation as
\begin{eqnarray} 
\ph^I(k,\eta) &=& \sum_{\lambda_1,\lambda_2} \ph^{\lambda_1\lambda_2}(k,\eta)
\delta^{\lambda_1\lambda_2}
= \sum_\lambda \ph^{\lambda\lambda}(k,\eta)\,.
\end{eqnarray}
Written explicitly in helical basis $\lambda=[L,R]$, 
$\ph^I = \ph^{LL} + \ph^{RR}\,.$
We may note here that only the even part of trispectrum ${\cal T}_{\rm even}$, 
contributes to $\ph^I$.
Note that the energy density of SIGW $\rho_{\rm GW}$ is related to $\ph^I$ 
(using the sub-Hubble behavior of $h^\lambda_\vk$) as
\begin{eqnarray}
\f{\d \braket{\rho_{\rm GW}}}{\d \ln k} &=& 
\f{\Mpl^2}{16} \, \left(\f{k^2}{a^2}\right)\, \ph^I(k,\eta)\,,
\label{eq:drho-dlnk}
\end{eqnarray}
with $a$ being the scale factor.


On the other hand, we may compute $\ph$ associated with the degree of circular 
polarization $V$ as
\begin{align}
\ph^V(k,\eta) &= \f{4\,k^3}{\pi^2\,(k\eta)^2}\iint \f{\d^3 \vq_1 \d^3 \vq_2}{(2\pi)^6}\,
[Q^{L}(\vk,\vq_1) Q^{L}(-\vk,-\vq_2) \nn \\
& - Q^{R}(\vk,\vq_1) Q^{R}(-\vk,-\vq_2)] \nn \\
& \times \tilde I_2\left(\f{q_1}{k},\f{\vert \vk - \vq_1\vert}{k},
\f{q_2}{k},\f{\vert \vk - \vq_2\vert}{k}\right) \nn \\
& \times \, {\cal T}_{\rm odd}(\vq_1,\vk-\vq_1,-\vq_2,\vq_2-\vk)\,.
\label{eq:phv}
\end{align}
By construction, $\ph^V$ is the difference between the power spectra of left and 
right helical modes $\ph^{LL} - \ph^{RR}$.
The factor $[Q^{L}(\vk,\vq_1) Q^{L}(-\vk,\vq_2) - Q^{R}(\vk,\vq_1) Q^{R}(-\vk,\vq_2)]$ 
in $\ph^V$ is imaginary and parity-odd, such that it extracts the contribution
from the parity-odd component of trispectrum ${\cal T}_{\rm odd}$.
Notice that if we consider individual helicities, ${\cal T}_{\rm odd}$ leads to an 
imbalance between the power in left and right helical modes, by diminishing 
one while amplifying the other by the same amount.
Apart from $\ps$ and $\cal T$, $\ph^I$ may receive contribution from scalar 
bispectrum~\cite{Unal:2018yaa,Cai:2018dig,Atal:2021jyo,Adshead:2021hnm,
Ragavendra:2021qdu,Perna:2024ehx,Yuan:2023ofl,Li:2023qua,Li:2023xtl}.
But $\ph^V$ shall remain insensitive to it as bispectrum preserves parity. 
As to the other Stokes parameters of this system, it can be shown that $Q=U=0$.

Similar to $\braket{\rho_{\rm GW}}$ we may relate the degree of circular polarization 
in the SIGW $V_{\rm GW}$, to $\ph^V$ as
\begin{eqnarray}
\f{\d \braket{V_{\rm GW}}}{\d \ln k} &=& \f{\Mpl^2}{16}\,\left(\f{k^2}{a^2}\right)
\,\ph^V(k,\eta)\,.
\label{eq:dv-dlnk}
\end{eqnarray}
The degree of chirality in $k$ space is then quantified by the ratio 
\begin{eqnarray}
\Pi_k &\equiv & \left(\f{\d \braket{V_{\rm GW}}}{\d \ln k}\right)\bigg/
\left(\f{\d \braket{\rho_{\rm GW}}}{\d \ln k}\right) \\
&=& \f{\ph^V(k,\eta)}{\ph^I(k,\eta)} = \f{\ph^{LL}(k,\eta) - \ph^{RR}(k,\eta)}{\ph^{LL}(k,\eta) + \ph^{RR}(k,\eta)}\,.
\end{eqnarray}
Since the time dependences are identical in both $\ph^I(k,\eta)$ and
$\ph^V(k,\eta)$, the ratio is time independent.
Besides, these time dependences are also cancelled by the prefactor
$(k/a)^2$ while relating them to $\rho_{\rm GW}$ and $V_{\rm GW}$ 
[cf.~Eqs.~\eqref{eq:drho-dlnk} and~\eqref{eq:dv-dlnk}].
So, for convenience in further usage, we define the combination
$\ph^{I,V}(k) = (k\eta)^2\ph^{I,V}(k,\eta)$\,.

Note that the Stokes parameters are constrained by the relation
$I^2 \geq Q^2 + U^2 + V^2$.
With $Q=U=0$, the relation implies $\vert V/I \vert \leq 1$. 

{\it Chirality induced}---For computing the chirality induced in SIGW, we
use parametric forms of the trispectrum $\cal T$ that capture the salient
features of it when generated by parity-violating inflationary 
models~\cite{Shiraishi:2016mok,Shiraishi:2016hjd,Niu:2022fki,Cabass:2022rhr,Fujita:2023inz,
Creque-Sarbinowski:2023wmb,Reinhard:2024evr,Moretti:2024fzb,Yura:2025mus,Philcox:2025bvj,Cho:2025rvg}.
The parity-odd part of the trispectrum shall be purely imaginary and contain 
vector triple products of the wavevectors involved to capture the change of 
signature under inversion of coordinates.
With these considerations, we examine two parametrizations of $\cal T$.

\underline{Template 1:}~We construct the odd and even parts of the first 
template of $\cal T$ as
\begin{subequations}
\begin{align}
{\cal T}_{\rm odd}(\vk_1,\vk_2,\vk_3,\vk_4) & = i\,\tgnl\, 
\beta(\widehat{\vk_1+\vk_2}, \hvk_1, \hvk_3)
P(k_1)P(k_3)\nn \\ & \times P(\vert\vk_1+\vk_2\vert) 
+\,\text{23 permutations}\,, \\
{\cal T}_{\rm even}(\vk_1,\vk_2,\vk_3,\vk_4) & = 2\,\gnl\, 
P(k_1)P(k_3)P(\vert\vk_1+\vk_2\vert) \nn \\ & 
+\,\text{11 permutations}\,.
\end{align}
\label{eq:trispec-paramet-1}
\end{subequations}
The overall amplitudes of ${\cal T}_{\rm odd}$ and ${\cal T}_{\rm even}$
are set by the dimensionless real parameters $\tgnl$ and $\gnl$ respectively.
The parity violating role in ${\cal T}_{\rm odd}$ is played by the function 
$\beta(\hvk_1, \hvk_2, \hvk_3) = \hvk_1 \cdot (\hvk_2 \times \hvk_3)\,,$
which under inversion ($\vk_i \to -\vk_i)$ acquires a negative sign.
The quantity $P(k)$ is the dimensionful scalar power spectrum, related to the 
dimensionless $\ps(k)$ as $P(k)=2\pi^2\,\ps(k)/k^3$\,.
Having the structure of $\cal T$ expressible in terms of $\ps$, the 
corresponding $\ph^V$ and $\ph^I$ can be computed, once the form of $\ps$ is 
specified.
We shall work with two cases of $\ps$: a scale-invariant form of $\ps(k)=\As$ and 
$\ps(k)$ containing a lognormal peak.

We shall begin with the case of $\ps(k)=\As$.
Rewriting ${\cal T}_{\rm odd}$ and  ${\cal T}_{\rm even}$ in terms of $\As$, $\gnl$
and $\tgnl$, we compute $\ph^V$ and $\ph^I$ numerically to obtain the measure of
chirality as
\begin{eqnarray}
\Pi_k &=& \f{\tgnl\,\As^3\,{\cal I}^{(2)}_V}{\As^2\,{\cal I}^{(1)}_I
+ \gnl\,\As^3\,{\cal I}^{(2)}_I}\,,
\end{eqnarray}
where ${\cal I}^{(1)}_I$, ${\cal I}^{(2)}_I$ and ${\cal I}^{(2)}_V$ are 
the numerical factors arising from the integrals over $\vq_1$ and $\vq_2$ in 
Eqs.~\eqref{eq:ph-l1-l2} and~\eqref{eq:phv}.
The factor ${\cal I}^{(1)}_I$ arises from performing the three-dimensional integral 
containing product of $\ps$ while ${\cal I}^{(2)}_I$ arises from the six-dimensional 
integral containing ${\cal T}_{\rm even}$ in Eq.~\eqref{eq:ph-l1-l2}, 
while computing $\ph^I=\sum_\lambda\ph^{\lambda\lambda}$.
The factor ${\cal I}^{(2)}_V$ arises from performing the six-dimensional integral 
containing ${\cal T}_{\rm odd}$ in Eq.~\eqref{eq:phv}.
We should note that, if we are to include contributions from bispectrum at the
same order in $\As$, parametrized by $\fnl$ with a template of choice (say local), 
then the expression of $\Pi_k$ shall extend to be
\begin{eqnarray}
\Pi_k &=& \f{\tgnl\,\As^3\,{\cal I}^{(2)}_V}{\As^2\,{\cal I}^{(1)}_I
+ \As^3\left(\fnl^2\,{\cal I}^{(3)}_I+ \gnl\,{\cal I}^{(2)}_I\right)}\,,
\label{eq:Pi_k-full}
\end{eqnarray}
where the additional factor of ${\cal I}_I^{(3)}$ shall arise from the
six-dimensional integral capturing the contribution from the 
bispectrum~\cite{Unal:2018yaa,Adshead:2021hnm,Ragavendra:2021qdu,Perna:2024ehx}.
But, since we are interested in the imprints due to the trispectrum, we shall
not delve into explicit computation related to bispectrum in our analysis.

Motivated by perturbativity, in this case we assume that the contribution from
bispectrum and ${\cal T}_{\rm even}$ to $\ph^I$ is sub-dominant to 
the contribution from $\ps^2$, i.e. 
$(\fnl^2\,{\cal I}^{(3)}_I,\,\gnl{\cal I}^{(2)}_I)\,\As < {\cal I}^{(1)}_I$~\cite{Garcia-Saenz:2022tzu}.
Hence
\begin{eqnarray}
\Pi_k &\simeq& \f{\tgnl\,\As^3\,{\cal I}^{(2)}_V}{\As^2\,{\cal I}^{(1)}_I}
\simeq -\f{7}{2\pi} \tgnl\,\As\,,
\end{eqnarray}
with the prefactor determined from numerical computation of the integrals
(see appendix for details of numerical integration).
Setting $\As = 2 \times 10^{-9}$ consistent with the normalization of 
$\ps$ over CMB scales~\cite{Planck:2018jri} and $\tgnl=1$, we obtain 
$\Pi_k \simeq -2 \times 10^{-9}\,.$
Since, we have assumed $\ps(k)=\As$ and constructed ${\cal T}_{\rm odd}$ in 
terms of it, we obtain scale-invariant $\ph^V$ and $\ph^I$.
The Stokes inequality $\vert V/I \vert \leq 1$ in this case becomes $\Pi_k \leq 1$,
which translates to model parameters as
\begin{equation}
\vert\tgnl\vert\As \lesssim 1\,.
\end{equation}
This limit mimics the bound due to perturbativity on $\gnl$, although it is on 
$\tgnl$ and stems from Stokes inequality.
It sets a bound of $\vert\tgnl\vert \lesssim 5\times 10^8\,.$
Note that this is a purely theoretical constraint and it does not rely on the exact amplitude
of tensor power spectra over CMB scales.
For comparison with observational bound on $\tgnl$, the closest available 
constraint is on a template of trispectrum arising from Chern-Simons 
coupling during inflation, which resembles Eq.~\eqref{eq:trispec-paramet-1}, 
barring factors involving scalar products of 
wavevectors~\cite{Shiraishi:2016mok,Philcox:2022hkh}.
The constraints on parameters equivalent to $\tgnl$ in this case is 
$\tgnl = 570 \pm 780$ from galaxy surveys~\cite{Philcox:2022hkh} and 
$\tgnl = (-1.4\pm 1.6)\times 10^4$ from CMB~\cite{Philcox:2025wts}.
Our constraint on $\tgnl$ is weaker by about three to four orders of magnitude.
However, we can apply the above arguments to scales smaller than CMB. There, 
if $\As$ reaches higher amplitudes and stays scale-invariant over a range of scales, 
the bound of $\tgnl \lesssim \As^{-1}$ shall become proportionately stronger.

We repeat the exercise for the case where $\ps$ has a lognormal peak and is enhanced 
over small scales. Such a case is relevant for generating SIGW of amplitudes 
detectable by current or upcoming observational missions, and also widely considered 
in the context of production of primordial black holes (PBH)~\cite{Bartolo:2016ami,Bartolo:2018evs,Bartolo:2018rku,LISACosmologyWorkingGroup:2023njw,
LISACosmologyWorkingGroup:2024hsc,LISACosmologyWorkingGroup:2025vdz,ET:2025xjr}. 
The form of $\ps$ in this case is parametrized as
\begin{align}
\ps(k) &= \As\left(\f{k}{k_\ast}\right)^{n_{\rm s} -1}
+ \f{A_{\rm p}}{\sqrt{2\pi\sigma^2_{\rm p}}}
\exp\left[-\f{1}{2\sigma_{\rm p}^2}
\ln^2\left( \f{k}{k_{\rm p}} \right)\right]\,.
\label{eq:ps-peak}
\end{align}
We set $\As = 2 \times 10^{-9}$ as before and $\ns = 0.96$
to be consistent with CMB over large scales.
We choose $A_{\rm p} = 10^{-2}$ so as to mimic models that generate significant 
amount of PBH and SIGW.
Such $\ps$, when plugged in to Eqs.~\eqref{eq:trispec-paramet-1}, 
in turn leads to a peaked trispectrum.
In this case, we account for the contribution from ${\cal T}_{\rm even}$ to $\ph$ 
along with the reducible term.
We set $\gnl = \tgnl = 1$ to numerically compute $\ph^I$ and $\ph^V$ and examine 
the chirality induced.

\begin{figure*}
\centering
\includegraphics[width=0.3\linewidth]{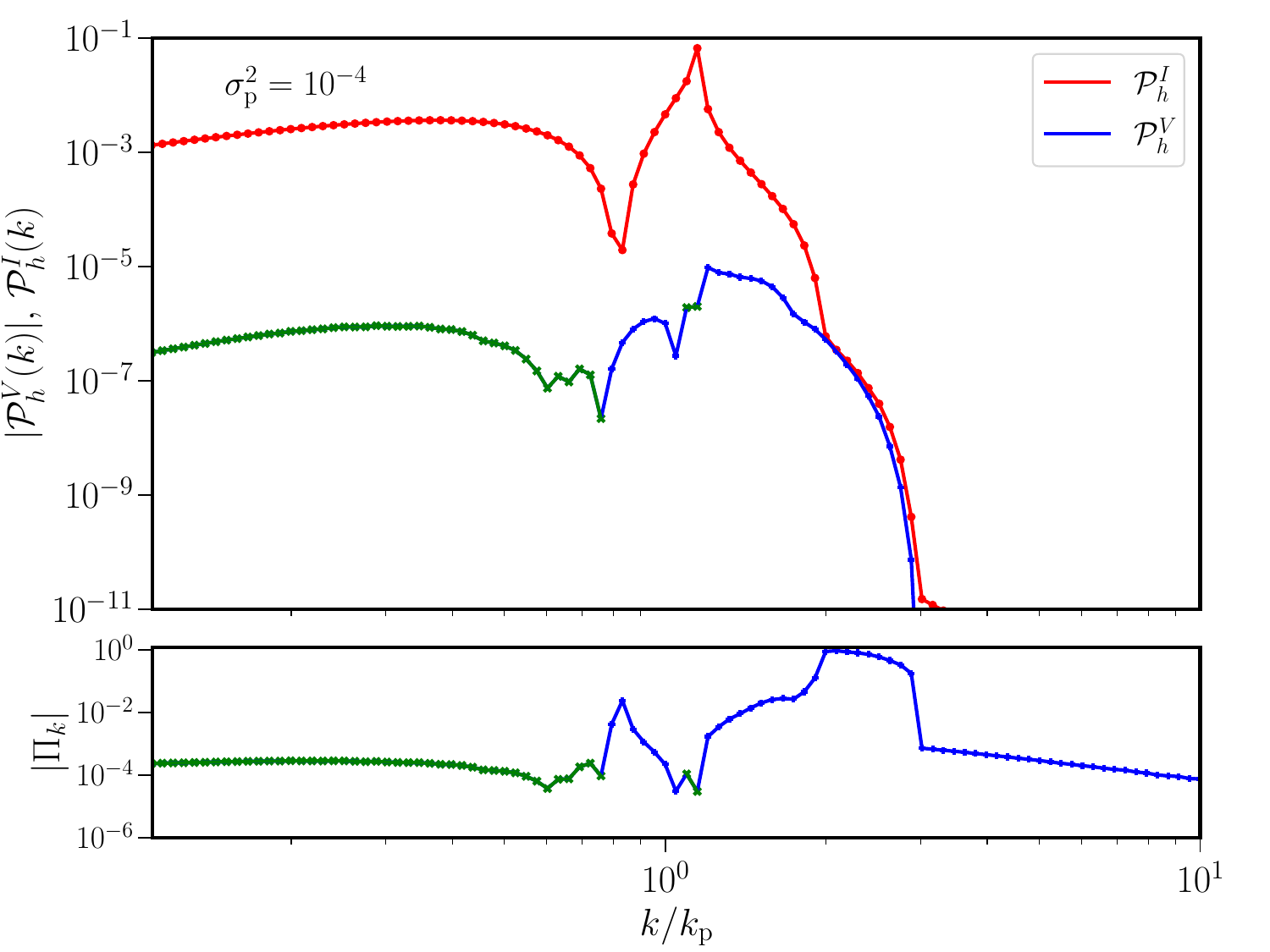}
\includegraphics[width=0.3\linewidth]{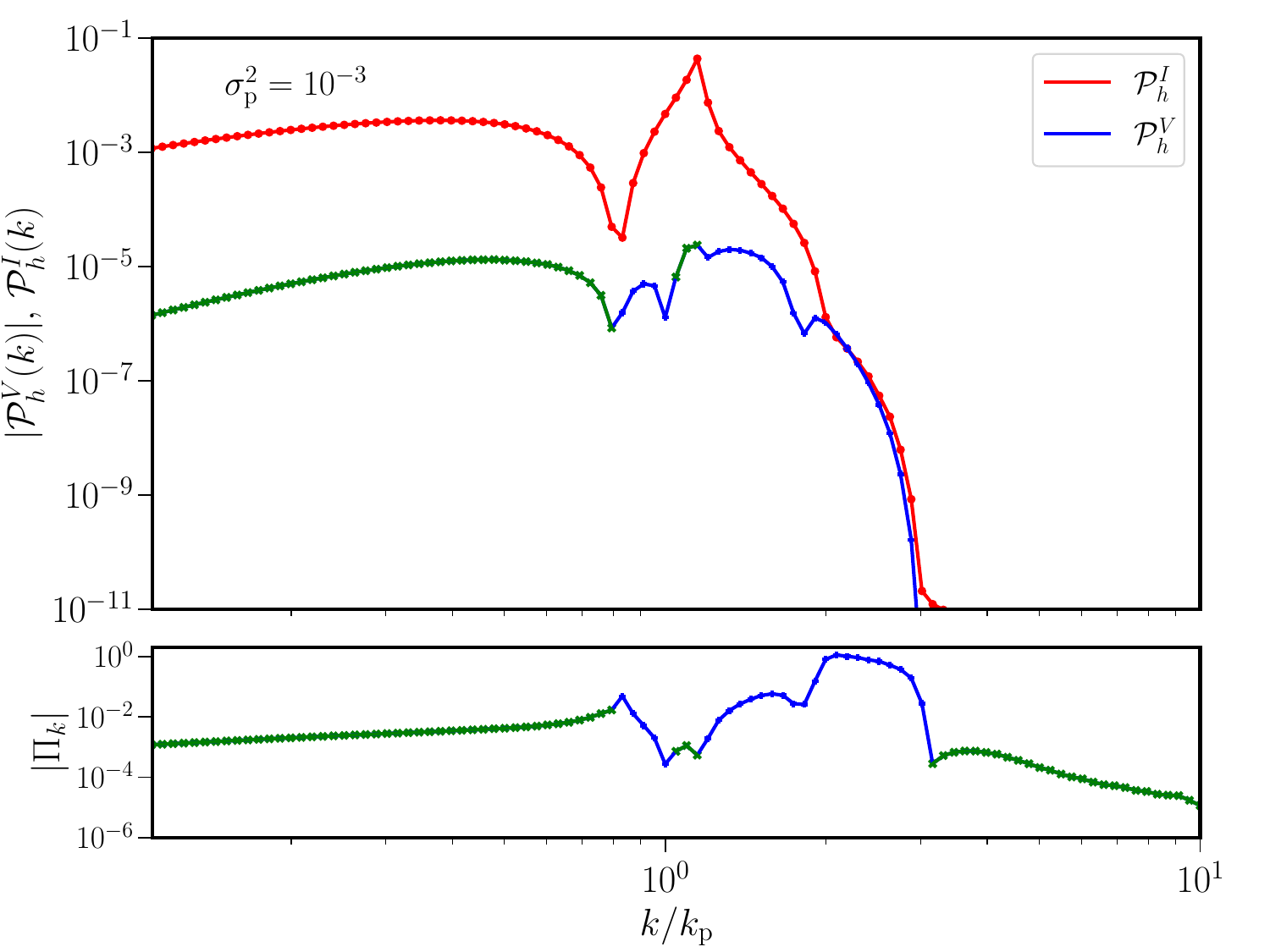}
\includegraphics[width=0.3\linewidth]{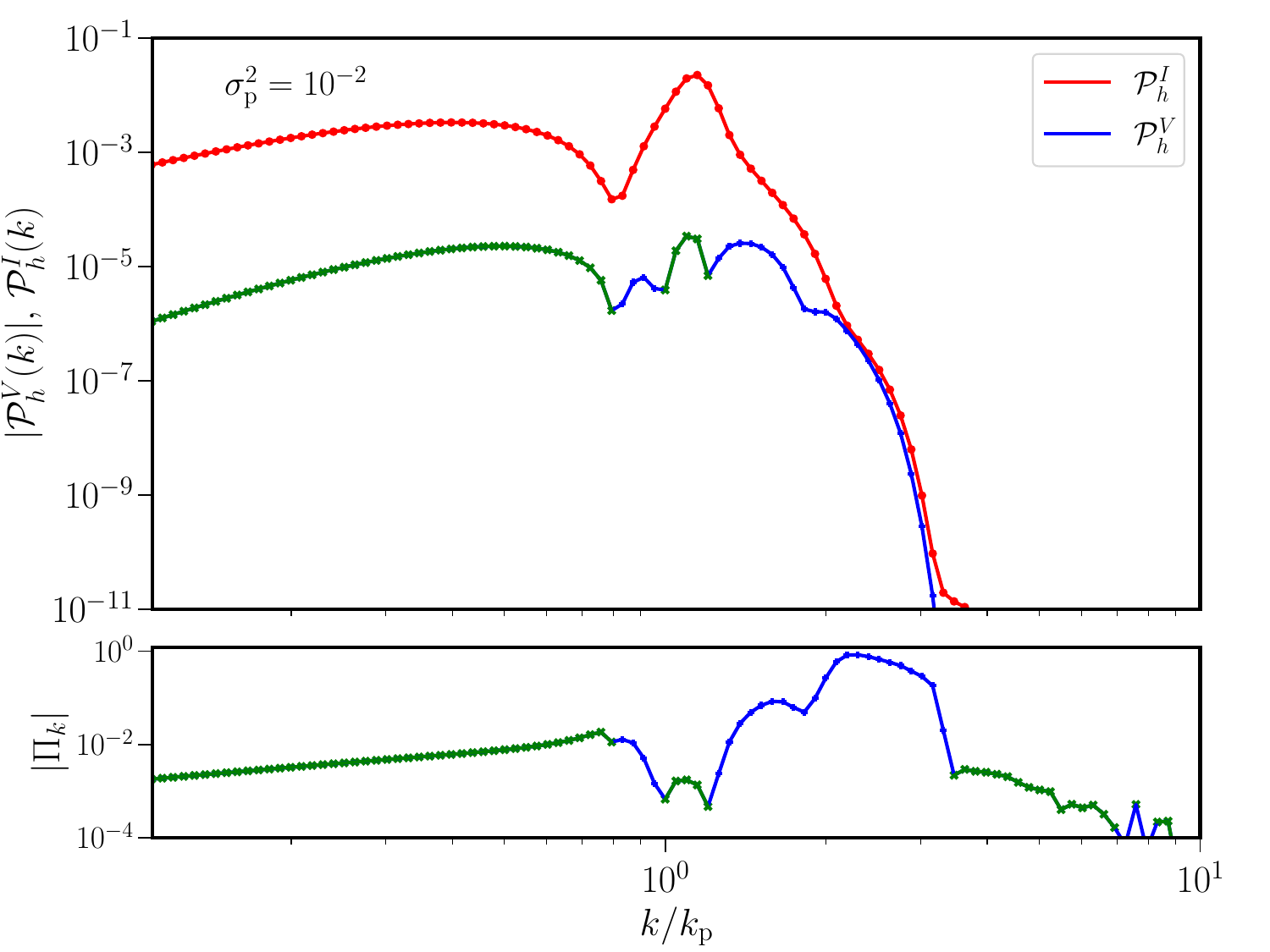}
\vskip -0.1in
\caption{We illustrate the spectral densities $\ph^V(k)$ and $\ph^I(k)$ 
computed for Template 1 of $\cal T$ [Eq.~\eqref{eq:trispec-paramet-1}], for 
the case of $\ps(k)$ with a lognormal peak [Eq.~\eqref{eq:ps-peak}], focussing 
over a range of scales close to the peak.
We present $\ph^I(k)$ and $\ph^V(k)$ in red and blue curves respectively, and
indicate negatives values of $\ph^V(k)$ in green.
We have set $A_{\rm p}=10^{-2}$, $\gnl=\tgnl=1$ and vary the width of the peak as 
$\sigma^2_{\rm p} = [10^{-4},10^{-3},10^{-2}]$ (as marked in each panel), 
inspecting the peak shift from a very sharp feature to a rather smooth feature 
in the spectrum. 
We also present the degree of chirality $\Pi_k$ at the bottom of each panel 
(positive values in blue and negative in green).
Note that $\ph^I(k)$ is computed including contributions from both $\ps$ and
${\cal T}_{\rm even}$.
We observe the following behaviors from this exercise:
$\Pi_k$ is highly scale-dependent, changing signature around the peak 
($k \simeq k_{\rm p}$), and of opposite sign compared to the region away 
from the peak (say $k \ll k_{\rm p}$).
$\Pi_k$ is considerably large over the wavenumbers post the peak at 
$k \simeq 3k_{\rm p}$, reaching as high as $\Pi_k \simeq 0.1-1$.}
\label{fig:ph-peak}
\end{figure*}
We present the behavior of $\ph^V(k)$ and $\ph^I(k)$ in this case along with 
their ratio $\Pi_k$ in Fig.~\ref{fig:ph-peak}.
We vary the width of the peak $\sigma_{\rm p}$ over a range of values to 
vary the feature from a sharp spike to a smooth curve and observe several 
interesting features.
Firstly, chirality alters sign around the peak, i.e. $\Pi_k$ is negative 
(right-circular) away from the peak over $k \ll k_{\rm p}$ and positive 
(left-circular) just after the peak $k \gtrsim k_{\rm p}$.
Secondly, $\tgnl=1$ easily induces chirality of at least $\Pi_k \simeq 10^{-3}$
around the peak. Importantly, at $k \simeq 3k_{\rm p}$, $\Pi_k$ reaches as high 
as $\Pi_k \simeq 0.1-1$, depending on the width of the peak.

Focussing specifically over this region, we find that the behavior of $\ph^I(k)$ 
around $k \simeq 3k_{\rm p}$ is dominated by contribution from 
${\cal T}_{\rm even}$~(and bispectrum, refer for instance~\cite{Perna:2024ehx}).  
Hence, the degree of chirality can be approximated as
\begin{equation}
\Pi_{k \simeq 3k_{\rm p}} \simeq \f{\tgnl\,{\cal I}^{(2)}_V}{\gnl\,{\cal I}^{(2)}_I}\,,
\label{eq:pi-tgnl-gnl}
\end{equation}
where ${\cal I}^{(2)}_{I,V}$ are numerical factors arising from the integrations
[cf.~Eqs.~\eqref{eq:ph-l1-l2} and~\eqref{eq:phv}], as mentioned before.
For trispectrum with similar shapes for parity-odd and parity-even components 
(barring $\beta$), ${\cal I}^{(2)}_I \simeq {\cal I}^{(2)}_V$, implying degree of chirality
to be $\Pi_{k \simeq 3k_{\rm p}} \simeq {\tgnl}/{\gnl}\,,$
the ratio of amplitudes of the parity-odd to parity-even parts of the trispectrum.
Examining Fig.~\ref{fig:ph-peak} in this range, $\Pi_k$ reaches substantial 
values informing us of the abovementioned ratio.
However, as noted earlier, above estimate does not include the contribution from
bispectrum $\As^3\fnl^2\,{\cal I}^{(3)}_I$ to $\ph^I$ [cf.~Eq.~\eqref{eq:Pi_k-full}].
If we compute this contribution, the ratio shall get modified as
\begin{equation}
\Pi_{k \simeq 3k_{\rm p}} \simeq \f{\tgnl\,{\cal I}^{(2)}_V}{\fnl^2{\cal I}^{(3)}_I +\gnl{\cal I}_I^{(2)}}\,,
\label{eq:pi-tgnl-gnl-fnl}
\end{equation}
with factors of ${\cal I}^{(3)}_I \simeq {\cal I}^{(2)}_I$ arising from similar integrals.
For values of $\fnl < \gnl$, $\Pi_{k \simeq 3 k_{\rm p}}$ reduces from Eq.~\eqref{eq:pi-tgnl-gnl} 
with corrections of order ${\fnl^2/\gnl}$.
When $\fnl^2=\gnl$, we see that $\Pi_{k\simeq 3 k_{\rm p}}$ is 
diminished roughly by a factor of $2$.
For $\fnl^2 > \gnl$, $\Pi_{k\simeq 3 k_{\rm p}} \simeq \tgnl/\fnl^2$.
Therefore, the chirality of SIGW, over a range of scales adjacent to the peak, 
provides a direct window to observe the relative amplitude of parity-violation 
present in primordial scalar non-Gaussianity.

\underline{Template 2}:~We consider another template of ${\cal T}$ to inspect 
the scenario where the trispectrum has an intrinsic scale-dependence, 
apart from the behavior of $\ps$.
Such a template is motivated by models of inflation that generate peaks in the 
spectra, often leading to non-Gaussianity parameters such as $\fnl$, having
non-trivial scale-dependences~\cite{Atal:2018neu,Atal:2019cdz,Ragavendra:2020sop,
Ragavendra:2021qdu,Ozsoy:2021pws,Pi:2022ysn,Ragavendra:2023ret,Fujita:2023inz}.
We model the odd and even parts of $\cal T$ as containing intrinsic peaks as
\begin{subequations}
\begin{align}
{\cal T}_{\rm odd}(\vk_1,\vk_2,\vk_3,\vk_4) & = i\,\tgnl
\bigg\{1 + \f{\alpha}{\sqrt{2\pi\sigma^2_t}}\nn \\ & 
\times \exp\left[\f{-1}{2\sigma^2_{\rm t}}\ln^2\left(\f{k_1+k_2+k_3+k_4}{4 k_{\rm t}}\right)\right]\bigg\} \nn \\ 
& \times \big[\beta(\widehat{\vk_1+\vk_2}, \hvk_1, \hvk_3)\,
P(k_1)P(k_3)\nn \\ & \times P(\vert\vk_1+\vk_2\vert) 
+\,\text{23 permutations}\big]\,, \\
{\cal T}_{\rm even}(\vk_1,\vk_2,\vk_3,\vk_4) & = 2\,\gnl\bigg\{1 + \f{\alpha}{\sqrt{2\pi\sigma^2_t}}\nn \\ &
\times \exp\left[\f{-1}{2\sigma^2_{\rm t}}\ln^2\left(\f{k_1+k_2+k_3+k_4}{4 k_{\rm t}}\right)\right]\bigg\} \nn \\ 
 & \times \big[P(k_1)P(k_3)P(\vert\vk_1+\vk_2\vert) \nn \\
 & +\,\text{11 permutations}\big]\,.
\end{align}
\label{eq:trispec-paramet-2}
\end{subequations}
The parameter $\alpha$ quantifies the relative enhancement of the overall amplitude
at the peak $k_{\rm t}$, compared to the value away from the peak.
We use $\ps(k)$ of the form Eq.~\eqref{eq:ps-peak} in this template and 
set $k_{\rm p}=k_{\rm t}$, as is often observed in inflationary models.
We vary the width of the peak in $\cal T$, $\sigma_t$, while fixing the value 
of $\sigma_p$ to inspect the interplay of scale-dependences between power and 
tri-spectra in the resultant behaviors of $\ph^V$ and $\ph^I$.

\begin{figure*}
\centering
\includegraphics[width=0.3\linewidth]{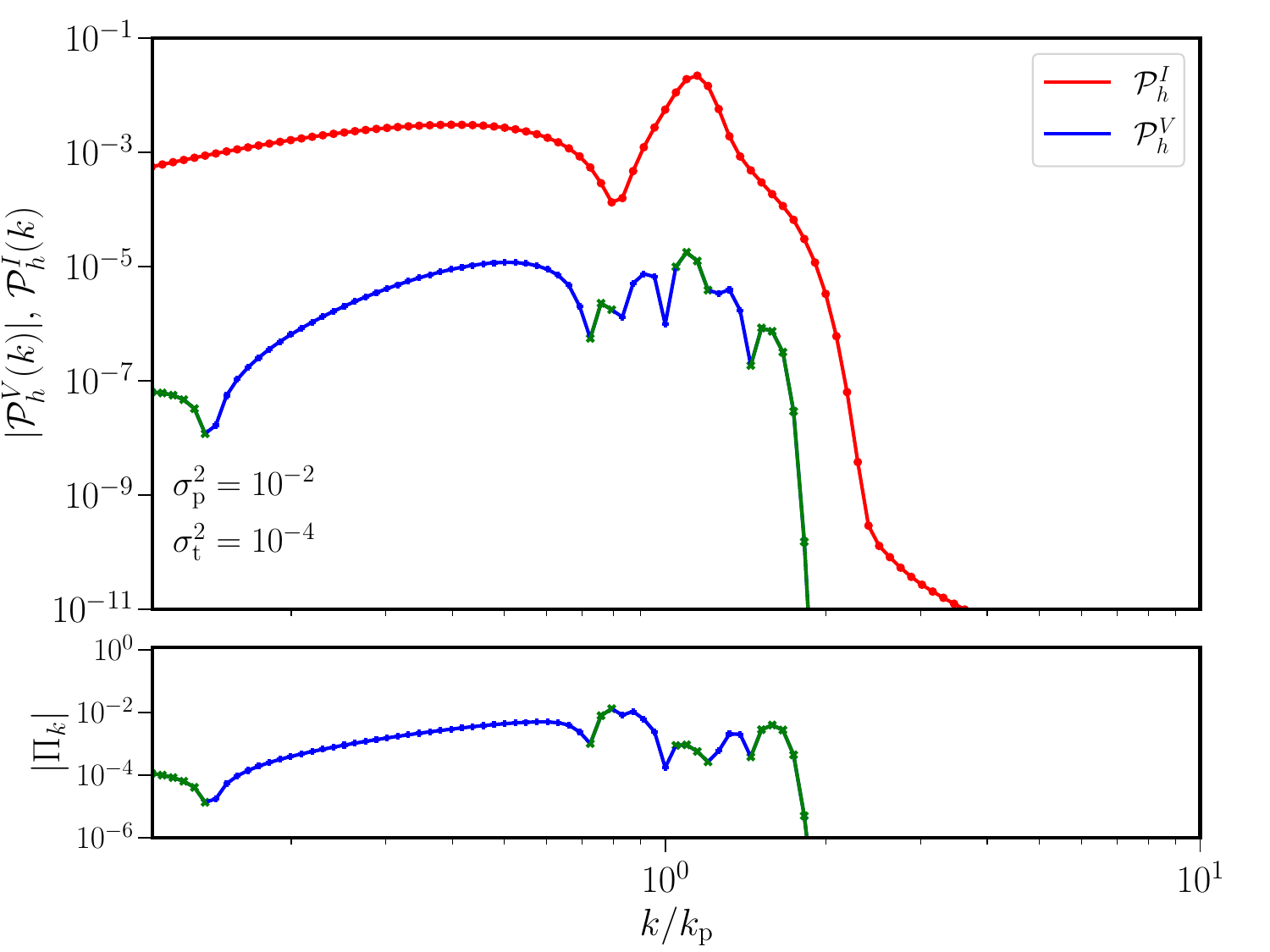}
\includegraphics[width=0.3\linewidth]{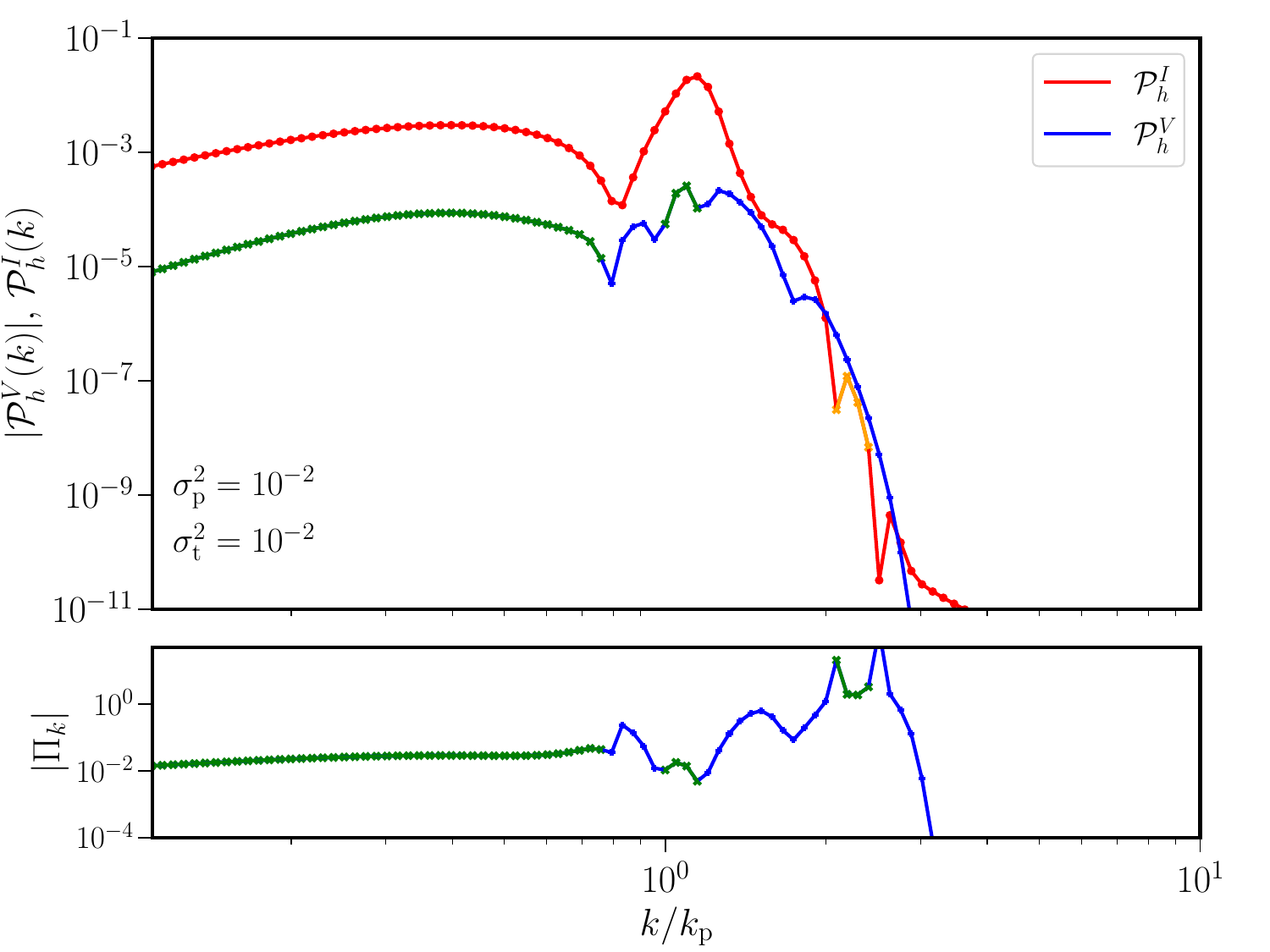}
\includegraphics[width=0.3\linewidth]{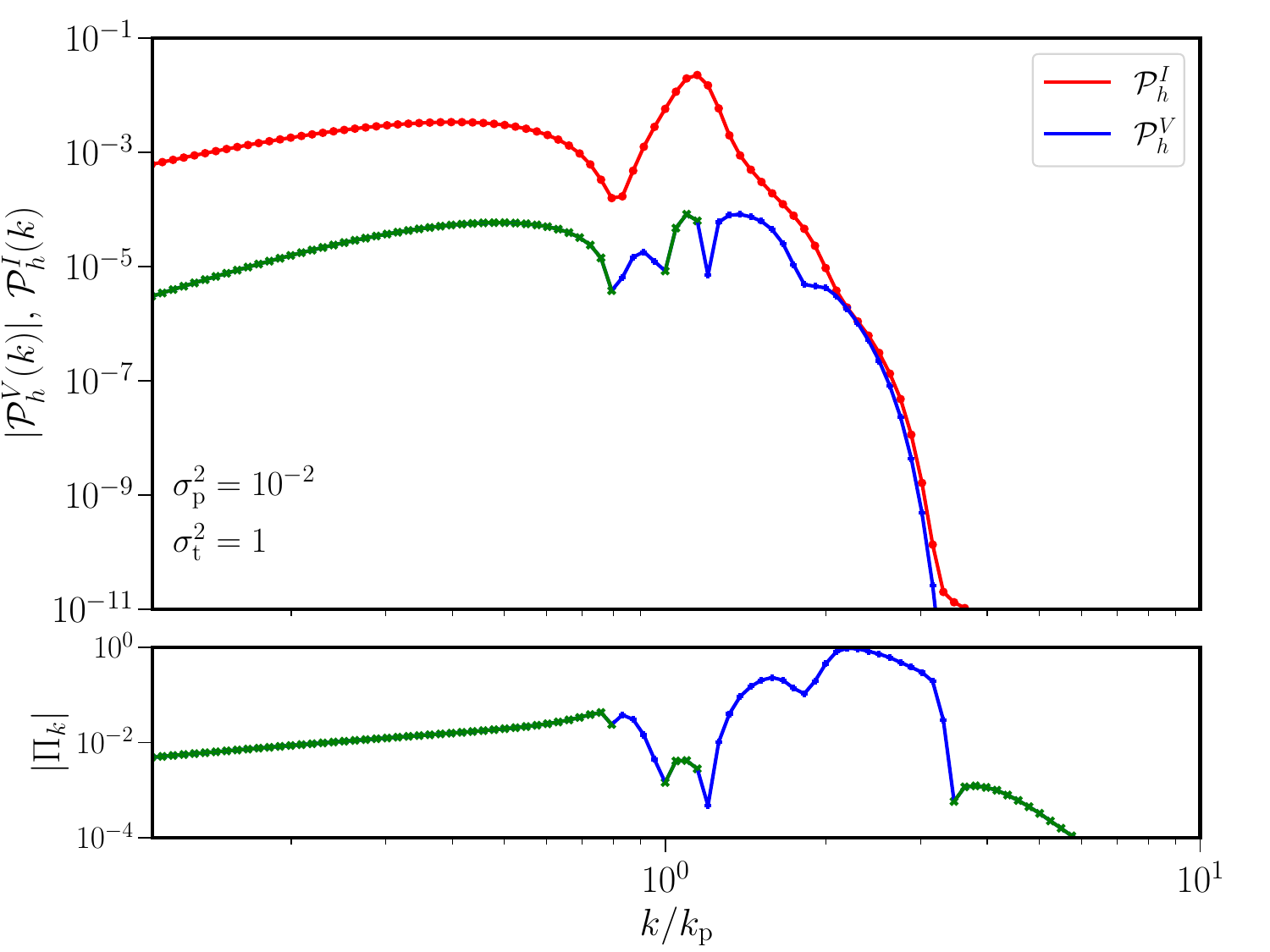}
\vskip -0.1in
\caption{We present $\ph^V(k)$ and $\ph^I(k)$, along with $\Pi_k$ (with same 
color choices as in Fig.~\ref{fig:ph-peak}) computed for Template 2 with $\ps(k)$ 
and $\cal T$ containing lognormal peaks 
[cf.~Eqs.~\eqref{eq:ps-peak} and~\eqref{eq:trispec-paramet-2}].
We also indicate the regimes where $\ph^V(k)$ and $\ph^I(k)$ turn negative 
(in green and orange respectively).
We set the parameters to be $\alpha=10$, $k_{\rm t}=k_{\rm p}$, 
$\sigma^2_{\rm p}=10^{-2}$ and others as mentioned in the text.
We vary $\sigma^2_{\rm t}=[10^{-4},10^{-2},1]$ to inspect the effect of such 
an intrinsic scale-dependence of $\cal T$ on $\ph^V(k)$ and $\ph^I(k)$.
We learn that for $\sigma^2_{\rm t}=10^{-4}$, the contribution from $\cal T$
to $\ph^I$ and $\ph^V$ is lesser and hence $\Pi_k\leq 10^{-2}$.
For $\sigma^2_{\rm t}=10^{-2}=\sigma^2_{\rm p}$, the contributions from $\cal T$
and $\ps$ to $\ph^I$ compete closely around $k_{\rm p}$, leading to corrugated
features in it. Further, $\ph^V$ becomes comparable and even dominant over 
$\ph^I$ in this brief window.
When $\sigma^2_{\rm t}=1$, the peak just rescales $\gnl$ and $\tgnl$ and
we recover behaviors of spectra similar to the case of $\sigma^2_{\rm p}=10^{-2}$ 
in Template 1 (cf.~Fig.~\ref{fig:ph-peak}).}
\label{fig:ph-peak-temp2}
\end{figure*}
We repeat the numerical analysis for this template and present the behaviors of 
$\ph^V$, $\ph^I$ and $\Pi_k$ around the peak in Fig.~\ref{fig:ph-peak-temp2}.
We observe that for $\sigma_{\rm t}^2 \ll \sigma_{\rm p}^2$, the contribution 
from $\cal T$ is highly suppressed and hence $\Pi_k \leq 10^{-2}$ around the peak, 
because of the contribution due to $\cal T$ being very much restricted over the
ranges of the integrations involved.
When $\sigma_{\rm t}^2=\sigma_{\rm p}^2$, we find that $\Pi_k$ is enhanced 
beyond $0.1$ around $k\simeq 3k_{\rm p}$. 
Curiously, $\ph^I$ is affected by features from the trispectrum and hence incurs 
a corrugated shape in this regime\footnote{The value of $\ph^I$ turning negative 
when dominated by ${\cal T}_{\rm even}$ 
indicates the necessity to account for other non-Gaussian contributions, 
such as the positive definite term proportional to $\fnl^2 A^3_{\rm p}$.}.
With $\sigma_{\rm t}^2 \gg \sigma_{\rm p}^2$, we find $\Pi_k$ recovering a behavior
close to Template 1 as in Fig.~\ref{fig:ph-peak}. 
This can be understood as the peak in $\cal T$ appearing wide enough for the peak in 
$\ps$, to seem like a constant rescaling of $\gnl$ and $\tgnl$. 
Further, inclusion of the contribution containing $\fnl^2 A^3_{\rm p}$ shall 
increase $\ph^I$ thereby reducing $\Pi_k$ as remarked in case of Template 1.
Besides the exact effect due to the peak in $\cal T$, this exercise underlines the
necessity to appropriately account for intrinsic scale-dependence in $\cal T$ while 
computing the spectra and chirality of SIGW, for it may lead to non-trivial
features.
Hence, a careful estimation of a scale-dependent $\Pi_k$ with GW surveys will lend
tighter constraint on $\tgnl$, along with $\gnl$ and $\As$ over corresponding scales. 
Such bounds over small scales shall complement the bounds on the parameters 
over large scales from CMB and galaxy surveys, thereby informing us of the extended
scale-dependence of primordial correlations.

{\it Conclusion}---In this work, we demonstrate that the parity-odd trispectrum 
leads to non-negligible chirality in SIGW.
We illustrate this phenomenon using two templates of $\cal T$, with two cases of
$\ps$ and numerically computing the resulting $\ph^V$ and $\ph^I$.
$\ph^V$ receives leading contribution from ${\cal T}_{\rm odd}$ and so is directly 
proportional to $\tgnl$.
The simple case of $\ps(k)=\As$, leads to a limit of $\tgnl \lesssim \As^{-1}$, 
sheerly due to the Stokes inequality $\vert V \vert \leq I$.
It translates to a conservative bound of $\vert \tgnl \vert \lesssim  5\times 10^8$
over large scales, consistent with the current observations~\cite{Planck:2019kim,
Philcox:2022hkh,Krolewski:2024paz,Philcox:2025wts}.
Our bound is purely theoretical and independent of the observability of GW
over the scales of interest.
Conversely, if $\tgnl$ is detected with sufficient confidence using CMB or galaxy 
surveys, our analysis provides a lower bound of $\Pi_k$, motivating the search 
for chirality in stochastic GW over corresponding frequencies.

For the case of $\ps(k)$ with a lognormal peak, we find that for $\tgnl=\gnl=1$
chirality can be as large as $\Pi_k \geq 10\%$ over scales adjacent to the peak.
Moreover, it alters in sign between scales close to and away from the peak.
We should emphasize that such values are well within the sensitivities of future 
observational missions.
For instance, the parameter choices of $A_{\rm p} \simeq 10^{-2}$ leads to 
$\ogw \simeq 10^{-10}$ at the peak~\cite{Kohri:2018awv,Domenech:2021ztg}.
If the peak occurs at small scales of $k_{\rm p} = {\cal O}(10^{15})\,\mpcinv$, it 
corresponds to frequencies of $f={\cal O}(1)\,{\rm Hz}$. 
Such $\ogw$ along with the V-mode $\ogw^V = \Pi_k\ogw$, can be probed using 
upcoming GW interferometers~\cite{Smith:2016jqs,Domcke:2019zls,Orlando:2020oko,Sato-Polito:2021efu,Cruz:2024esk}.
Note that $\Pi_k \geq 10\%$ over scales adjacent to the peak for nominal values
of $\tgnl=\gnl=1$.
When contributions in terms of $\fnl$ are included, the resultant $\ogw$ and the
associated $\ogw^V$, should be able to probe parameters $\fnl,\tgnl,\gnl$ of 
values ${\cal O}(10)$.
Hence, the bounds that we may obtain on $\tgnl$ over small scales through 
potential measurement of $\Pi_k$, with distinct scale-dependent signatures,
can be comparable to or even stronger than the limits obtained from CMB or galaxy 
surveys.
Such a window to directly assess the magnitude and signature of ${\cal T}_{\rm odd}$ 
is observationally rare and possibly unique over small scales.

With the Template 2, we illustrate the importance of scale-dependence of
$\cal T$ for the behavior of $\ph^V$ and $\ph^I$.
Crucially, using peaked spectra in both templates, we find a window of scales 
where $\Pi_k$ turns out to be the ratio of parity-odd to parity-even 
parts of non-Gaussian contributions to SIGW. Hence, $\Pi_k$ therein quantifies 
the parity-violation in the higher-order correlations of primordial scalar 
perturbations.

On the theoretical front, our results call for careful analyses of SIGW from 
parity-violating inflationary models.
As mentioned earlier, in models with suppressed chirality of primary GW, 
${\cal T}_{\rm odd}$ through SIGW shall determine the overall behavior of $\Pi_k$.
In models with maximally chiral primary GW, the total $\Pi_k$ shall be attenuated by the
contribution from SIGW.
Hence in general, ${\cal T}_{\rm odd}$ shall strongly impact the magnitude,
sign and scale-dependence of $\Pi_k$.
In essence, the predictions of ${\cal T}_{\rm odd}$ and $\Pi_k$ arising from 
parity-violating models may be treated as twin predictions, informing about each
other and alleviating observational degeneracies.

{\it Acknowledgments}---HVR and NB thank Ilaria Caporali, Sabino Matarrese, 
Tommaso Moretti, Gabriele Perna and Angelo Ricciardone for insightful comments.
The authors acknowledge support by the MUR PRIN2022 Project ``BROWSEPOL: Beyond 
standaRd mOdel With coSmic microwavE background POLarization''-2022EJNZ53 financed 
by the European Union - Next Generation EU. 
The authors acknowledge the use of computing cluster under 
University of Padova Strategic Research Infrastructure Grant 2017:
``CAPRI: Calcolo ad Alte Prestazioni per la Ricerca e l’Innovazion''.
The authors acknowledge support by the INFN InDark Initiative. 
NB acknowledges support by the COSMOS network (www.cosmosnet.it) though the ASI 
(Italian Space Agency) Grants 2016-24-H.0, 2016-24-H.1-2018 and 2020-9-HH.0. 
The authors acknowledge that part of this work was carried out during the 2025 
``The Dawn of Gravitational Wave Cosmology'' workshop, supported by the Fundacion 
Ramon Areces and hosted by the ``Centro de Ciencias de Benasque Pedro Pascual'',
and thank them for hospitality. 


\section{Appendix}


{\it Details of integration required for $\ph^I$ and $\ph^V$}--The factor 
$Q^\lambda$ in Eq.~\eqref{eq:h1h2} is a dimensionless function that arises from
the contraction of the polarization tensor (for a given polarization $\lambda$)
defined with respect to the propagation direction of $\hat \vk$, and the wavevector 
of integration $\vq$ in the following fashion
\begin{equation}
Q^\lambda(\vk,\vq) = e^\lambda_{ij}(\vk)\f{q_iq_j}{k^2}\,.
\end{equation}
Here $e^\lambda_{ij}(\vk)$ is the polarization tensor that can be written in
terms of the polarization vectors orthogonal to $\hat \vk$, in any basis,
say $\lambda=[+,\times]$ or $\lambda=[R,L]$ (refer~\cite{Domenech:2021ztg} and 
references therein for explicit constructions).

The function $\tilde I_2$ that appears in Eq.~\eqref{eq:h1h2} is a dimensionless
function arising from the evolution of the inhomogeneous solution of 
$h_\vk^\lambda$ (which is denoted as $\tilde I(kv,ku,k\eta)$ 
in~\cite{Adshead:2021hnm,Perna:2024ehx}).
It results from averaging over fast oscillations of the solution and the overall 
time dependence is extracted out as $1/(2k^2\eta^2)$.
The explicit expression of the function $\tilde I_2$ is given by~\cite{Adshead:2021hnm}
\begin{align}
\tilde I_2\left(u_1,v_1,u_2,v_2\right) &=
I_A(u_1,v_1)I_A(u_2,v_2) \bigg[I_B(u_1,v_1)I_B(u_2,v_2) \nn \\
& + \pi^2 I_C(u_1,v_1)I_C(u_2,v_2) \bigg]\,,
\end{align}
where
\begin{align}
I_A(u,v) &= \f{3(u^2+v^2-3)}{4u^3v^3}\,, \\
I_B(u,v) &= -4uv + (u^2+v^2-3)\ln \left\vert \f{3 - (u+v)^2}{3 - (u-v)^2}\right\vert\,, \\
I_C(u,v) &= (u^2+v^2-3)\Theta(u+v-3)\,.
\end{align}
From the expression above, we may obtain $\tilde I_1$ that appears in Eq.~\eqref{eq:ph-l1-l2} as
\begin{eqnarray}
\tilde I_1(u,v) &=& \tilde I_2(u,v,u,v)\,, \\
&=& I_A^2(u,v)\left[I_B^2(u,v) + \pi^2 I_C^2(u,v) \right]\,.
\end{eqnarray}

For numerical implementation, we perform the integrations in Eqs.~\eqref{eq:ph-l1-l2} 
and~\eqref{eq:phv} to compute $\ph^I$ and $\ph^V$ 
using Monte-Carlo method of importance sampling with $10^9$ samples for each 
$k$ value. 
Specifically, for the two- and five-dimensional integrals (with one angular 
integration always performed trivially due to homogeneity and isotropy) that 
correspond to terms with $\ps^2$ and $\cal T$ in the integrand, we perform 
uniform sampling over the ranges of polar angles as $\cos \theta_{1,2} \in [-1,1]$ 
and the combination of azimuthal angles $(\phi_1-\phi_2) \in [0,2\pi]$.
As to the ranges of $q_{1,2}$, we perform importance sampling with 
a Gaussian sampler in logarithmic range.
In case of the spectra without peaks, we center the sampler at 
$q_{1,2}=k$, i.e. $\log_{10}(q_{1,2}/k)=0$ with its standard deviation set 
to $5$, i.e. five decades of wavenumbers around the center.
In case of peaks in $\ps$ (and in $\cal T$ in Template 2), we center the sampler at the location of the peak as
$q_{1,2}=k_{\rm p}$, i.e. $\log_{10}(q_{1,2}/k)=\log_{10}(k_{\rm p}/k)$, 
with the standard deviation set to $5\sigma_{\rm p}$.
The center and standard deviation of the sampler are chosen such that the sampling covers the prominent volume of the integrand in each case with sufficient convergence.
The convergence for a given $k$ is checked by computing the sampling variance 
and ensuring the ratio of square-root of this variance to the integral to be 
$\lesssim 10\%$.

Regarding the behavior of the integrand in case of scale-invariant spectra, 
there is a logarithmic dependence of the integrand in the form of $[\log(q_{1}/k)]^2$ 
specifically when $q_{1}/k \to \infty$ with $q_2=k$ and vice versa ($q_1 \leftrightarrow q_2 $).
On the other hand, certain terms in the trispectrum contain 
$\vert \vq_1 - \vq_2\vert^{-3}$ and diverge when $\vq_1 \to \vq_2$.
Through change of variables, it can be shown that it is same as the divergence 
of $q_{1,2}^{-3}$ when $q_{1,2}\to0$.
For numerical implementation, we have chosen a finite minimum value of integrated wavenumbers as 
$q_{1,2}/k = 10^{-3}$ and have checked that the results do not strongly depend on this choice.
We also note that these divergences are highly suppressed in case of peaked power and tri-spectra.

We should note that $\ph^V$ is sensitive to the three-dimensional nature of the 
trispectrum, as ${\cal T}_{\rm odd}$ vanishes if any of the constituting wavevectors 
is taken to zero, or equivalently any two of them are set to be (anti-)parallel.
Besides, not all configurations of ${\cal T}_{\rm odd(even)}$ contributes to 
$\ph^{V(I)}$.
Terms of ${\cal T}$ that are insensitive to angles between wavevectors of 
integration $\vq_1$ and $\vq_2$, vanish upon integration over azimuthal angles.
Also, of the different permutations of terms in $\cal T$ 
[cf.~Eqs.~\eqref{eq:trispec-paramet-1} and~\eqref{eq:trispec-paramet-2}], 
when substituted in the integrals of Eq.~\eqref{eq:ph-l1-l2}, some of them can 
be equated to others through change of variables, thereby reducing the unique 
types of terms that have to be integrated.

\bibliographystyle{apsrev4-2}
\bibliography{lib}

\end{document}